\documentclass[conference]{IEEEtran}
\IEEEoverridecommandlockouts
\usepackage{cite}
\usepackage{amsmath,amssymb,amsfonts}
\usepackage{algorithmic}
\usepackage{graphicx}
\usepackage{textcomp}
\usepackage{url}
\usepackage{numprint}
\usepackage{comment}
\usepackage[table]{xcolor}

\usepackage{booktabs}
\usepackage[caption=false,font=footnotesize,labelfont=sf,textfont=sf]{subfig}
\usepackage{tikz}
\usepackage{multirow}
\def\BibTeX{{\rm B\kern-.05em{\sc i\kern-.025em b}\kern-.08em
    T\kern-.1667em\lower.7ex\hbox{E}\kern-.125emX}}
\begin{document}

\title{SEF-MK: Speaker-Embedding-Free Voice Anonymization through Multi-k-means Quantization}

\author{
\begin{tabular}{c c}
\begin{tabular}{c}
Beilong Tang \\
Duke Kunshan University, China \\
{\tt\small beilong.tang@dukekunshan.edu.cn}
\end{tabular}
&
\begin{tabular}{c}
Xiaoxiao Miao\thanks{Xiaoxiao Miao is the corresponding author. This research is funded by DKU foundation project ``Emerging AI Technologies for Natural Language Processing'', and partially supported by JST, PRESTO Grant Number JPMJPR23P9. Many thanks for the computational resource provided by the Advanced Computing East China Sub-Center.} \\
Duke Kunshan University, China \\
{\tt\small xiaoxiao.miao@dukekunshan.edu.cn}
\end{tabular}
\\[2ex]
\begin{tabular}{c}
Xin Wang \\
National Institute of Informatics, Japan \\
{\tt\small wangxin@nii.ac.jp}
\end{tabular}
&
\begin{tabular}{c}
Ming Li \\
Duke Kunshan University, China \\
{\tt\small ming.li369@dukekunshan.edu.cn}
\end{tabular}
\end{tabular}
}

\maketitle

\begin{abstract}

Voice anonymization protects speaker privacy by concealing identity while preserving linguistic and paralinguistic content. Self-supervised learning (SSL) representations encode linguistic features but preserve speaker traits. We propose a novel speaker-embedding-free framework called SEF-MK. Instead of using a single k-means model trained on the entire dataset, SEF-MK anonymizes SSL representations for each utterance by randomly selecting one of multiple k-means models, each trained on a different subset of speakers. We explore this approach from both attacker and user perspectives. Extensive experiments show that, compared to a single k-means model, SEF-MK with multiple k-means models better preserves linguistic and emotional content from the user’s viewpoint. However, from the attacker’s perspective, utilizing multiple k-means models boosts the effectiveness of privacy attacks. These insights can aid users in designing voice anonymization systems to mitigate attacker threats. \footnote{Code and audio samples can be found at \url{https://github.com/Beilong-Tang/sef-mk}}

\end{abstract}


\section{Introduction}

Voice-based human-computer interaction is becoming increasingly prevalent, offering significant convenience in our daily lives. However, uploading raw audio recordings to social media without proper protection may lead to the leakage of personally identifiable information~\cite{GDPR}.
One form of additional protection is to encrypt utterances, thereby restricting access to unintended recipients and allowing only authorized users to recover the speech using a decryption key. However, this approach effectively prevents open communication and is not suitable for public-oriented applications. In most cases, users wish to convey the content and emotional tone of their speech while concealing their identity.
This is where a Voice Anonymization System (VAS) has been proposed~\cite{tomashenko2021voiceprivacy, tomashenko2022voiceprivacy, 10603395, tomashenko2024voiceprivacy}. Specifically, VAS processes the original speech to produce anonymized outputs in which speaker-identifiable information is removed (privacy), while essential attributes, such as linguistic content and emotional expression, are preserved to enable the use of anonymized speech for various downstream tasks (utility). The anonymized speech can be shared openly, mitigating concerns over the misuse of the source speaker's identity.

There are two mainstream approaches to voice anonymization.
Digital Signal Processing (DSP)-based methods are training-free approaches that modify speech characteristics from a speech production perspective to conceal the speaker’s identity. Techniques include altering formants~\cite{gupta2020design,dubagunta2020adjustable,mcadams1984spectral}, changing speech speed~\cite{vpc22tsm}, and modifying other vocal tract or voice source features~\cite{tavi2022improving}. However, DSP-based methods often suffer from content distortion and are generally ineffective against stronger attackers~\cite{tomashenko2020introducing, tomashenko2021voiceprivacy, srivastava2020evaluating}.

Deep Neural Network (DNN)-based methods, on the other hand, are generally more effective and draw on techniques from neural voice conversion and speech synthesis. 
A straightforward solution involves first transcribing speech into text using an automatic speech recognition (ASR) system, followed by re-synthesizing the speech via a text-to-speech (TTS) model~\cite{xinyuan24_spsc}.   
While this ASR+TTS pipeline has been shown to effectively conceal the speaker’s identity, it often compromises the utility of the original speech. Specifically, inevitable transcription errors and the loss of important paralinguistic cues, such as emotion, prosody, and accent, can degrade the quality and expressiveness of the anonymized output.

Widely used DNN-based approaches primarily leverage disentangled representation learning and rely on explicit speaker modeling, typically using pretrained speaker encoders. These approaches generally consist of three stages:
(i) Speech disentanglement aims to separate speaker-specific information from the input speech. Some methods explicitly disentangle speaker, content, and prosody components~\cite{fang2019speaker, meyer2023prosody, shamsabadi2022differentially, mawalim2022speaker, champion2023anonymizing, yao2024musa, miao22_odyssey, miao2023language, yao2024distinctive, meyer2024probing, miao2024benchmark}, often extracting content information with self-supervised learning (SSL) models~\cite{hsu2021hubert, chen2022wavlm, baevski2020wav2vec} and encoding speaker identity through a pretrained speaker encoder~\cite{snyder2018x, desplanques2020ecapa}. Others, inspired by speech codec technology, segment speech into acoustic and semantic tokens~\cite{panariello_speaker_2023}, treating acoustic tokens as the primary carriers of speaker-specific characteristics.
(ii) Speaker embedding anonymization involves replacing or transforming speaker-specific features with those of a pseudo speaker. Most approaches rely on an external speaker pool to construct the pseudo speaker~\cite{fang2019speaker, Srivastava2020DesignCF, srivastava2022privacy, yao2024distinctive,miao2022analyzing}. 
(iii) Speech generation~\cite{kong2020hifi} synthesizes anonymized speech by combining the anonymized speaker (or acoustic) features with the preserved content and prosody (or semantic) information. %This multi-stage process heavily relies on pretrained speaker encoders.

An emerging direction in DNN-based VAS research focuses on speaker-embedding-free neural methods~\cite{xinyuan24_spsc, cai2024privacy, ghosh2024anonymising, das2024comparing}, which avoid the explicit modeling of speaker embeddings. In these approaches, utterances from both source and pseudo speakers typically pass through a shared encoding stage that extracts speech representations using SSL models. Anonymization is then achieved by replacing the SSL feature frames from the source speaker with those from the pseudo speaker’s representation selected according to specific strategies, e.g. $k$-Nearest Neighbor (KNN)~\cite{baas2023voice}. The resulting representations are then used to synthesize anonymized speech.  However, since the SSL features contain speaker-related information, the $k$ nearest neighbors and the anonymized waveform may also encode the traits of the source speaker. These systems have been shown to provide poor speaker privacy protection when facing strong attacking~\cite{das2024comparing, xinyuan24_spsc}.

This paper explores the \textbf{s}peaker-\textbf{e}mbedding-\textbf{f}ree VAS paradigm and anonymizes SSL representations through \textbf{m}ultiple \textbf{k}-means quantization, referred to as \textbf{SEF-MK}. The method comprises three key stages:
(i) Encoding, where an SSL model-WavLM\cite{chen2022wavlm} generates continuous speech representations that capture linguistic content and (undesirably) speaker identity;
(ii) Multi-k-means quantization, where SSL representations are quantized to suppress speaker-specific traits by randomly selecting a k-means model from multiple quantizers; 
(iii) Decoding, where high-quality anonymized speech is reconstructed using a Conformer-based decoder~\cite{gulati2020conformer} and a HiFi-GAN vocoder~\cite{kong2020hifi}, ensuring naturalness while preserving privacy.

Prior work, such as KNN-based VAS~\cite{das2024comparing,xinyuan24_spsc}, has explored speaker-embedding-free approaches, none have employed k-means in this context, despite its demonstrated effectiveness in suppressing speaker identity~\cite{miao2023language}. Although k-means has been applied in disentanglement-based methods~\cite{miao2023language}, most studies use a single k-means model trained on a broad speaker population and primarily examine how varying the number of centroids affects the privacy-utility trade-off, lacking a thorough investigation of the specific role and effectiveness of k-means in anonymization.

This motivated us to pose two research questions, one from the user's perspective and the other from the attacker's, within the context of the VoicePrivacy Challenge (VPC)~\cite{tomashenko2024voiceprivacy}, which simulates a game-theoretic scenario between users and attackers. In this setting, users apply anonymization techniques to conceal speaker identities before publication, while attackers attempt to recover the original identities from the anonymized data. From the user's perspective, we explore strategies for training k-means models within the proposed SSL-based, speaker-embedding-free VAS paradigm by investigating the following question:
\noindent
\textit{Does the composition of the k-means training data affect anonymization performance?}
We construct a pool of k-means quantizers, each trained on a distinct subset of speakers using different speaker grouping strategies, such as training one model per individual speaker or training models on subsets of multiple speakers. We then investigate how these strategies affect anonymization performance from the user/defender's  perspective.
Through extensive experiments, we found that compared to a single k-means model, from the user's perspective, the use of multiple k-means models in the proposed \textbf{SEF-MK} system better preserves linguistic and emotional attributes. 

From the attacker's perspective, we investigate \emph{how the attacker's composition of the k-means degrade the anonymization performance}. We found that employing multiple k-means models can enhance the effectiveness of privacy attacks, regardless of which k-means strategy the user adopts. We hope these findings can be helpful for users when designing the VAS against attackers with various attacking strategies.

%==========================================
\begin{figure*}[h]
\centering
\includegraphics[width=1\textwidth]{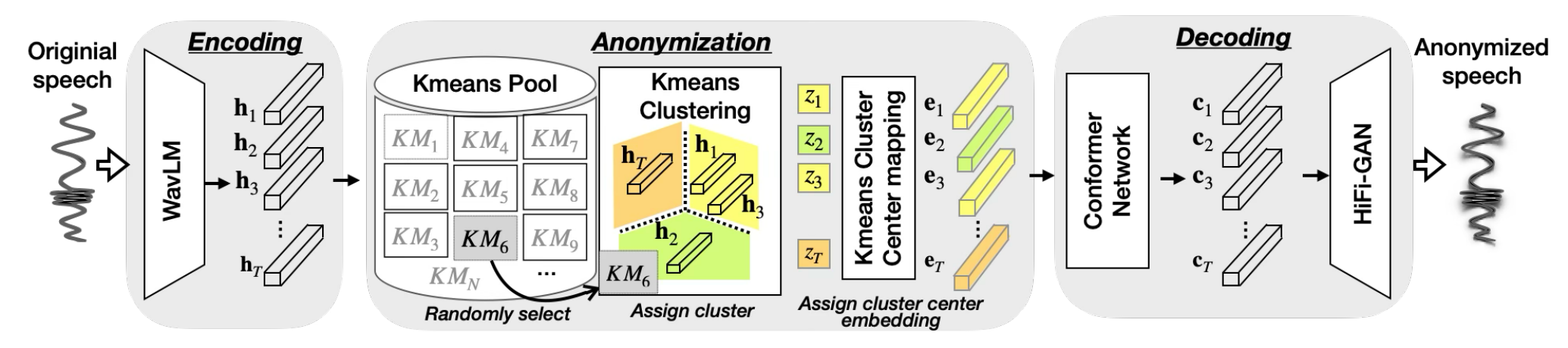}
\vspace{-20pt}
\caption{ Framework of multiple k-means-based speaker embedding-free voice anonymization.}
\label{fig:overall}
\vspace{-10pt}
\end{figure*}   
%==========================================

\section{Speaker-Embedding-Free Voice Anonymization}
In this section, we first provide an overview of the voice anonymization problem formulation and introduce two recently proposed speaker-embedding-free methods, which serve as baselines for this study. We then present the novel \textbf{SEF-MK} architecture, which leverages WavLM, k-means, Conformer, and HiFi-GAN to explore various strategies for training multiple k-means models to conceal the original speaker identity.

\subsection{Voice Anonymization Problem Formulation} 

An anonymization system transforms an original speech utterance $\mathbf{X} = [x_1, \ldots, x_{\tilde{T}}]$ into an anonymized version $\mathbf{Y} = [y_1, \ldots, y_{\tilde{T}}]$, where $\tilde{T}$ is the length of the utterance. An ideal anonymization system should preserve linguistic content and emotional cues\footnote{While paralinguistic attributes include not only emotion, this study follows the setup of Voice Privacy Challenge 2024~\cite{tomashenko2024voiceprivacy} and only considers emotion.} while removing the speaker identity. This objective can be formally defined as:

\vspace{-5pt}
\begin{equation}
\mathbf{Y} = f(\mathbf{X}),\quad \text{s.t.} 
\begin{cases}
\texttt{ASV}(\mathbf{X}) \neq \texttt{ASV}(\mathbf{Y}) \\
\texttt{ASR}(\mathbf{X}) \approx \texttt{ASR}(\mathbf{Y}) \\
\texttt{SER}(\mathbf{X}) \approx \texttt{SER}(\mathbf{Y})
\end{cases}
\end{equation}
The transformation function $f(\cdot)$ should simultaneously satisfy three competing objectives: (i) distort speaker identity, as measured by automatic speaker verification ($\texttt{ASV}$) systems, (ii) preserve linguistic content, as evaluated by an automatic speech recognition ($\texttt{ASR}$) system, and (iii) maintain emotional content, as assessed by a speech emotion recognition ($\texttt{SER}$) systems.

\subsection{KNN-based VAS}
\label{sec:knn}
Among speaker-embedding-free approaches, one popular approach is based on KNN~\cite{baas2023voice} with encoding, anonymization, and decoding stages.

In the encoding stage, both the source and target/pseudo speaker utterances go through the same pretrained WavLM model~\cite{chen2022wavlm} to extract SSL representations. 
In the anonymization stage, the goal is to remove speaker-specific information from the source WavLM representation while preserving other speech attributes. This is achieved by replacing each frame of the source representation with the most similar frames from the target speaker. Specifically, for each frame in the source utterance, a KNN search is performed to identify the top-$k$ most similar frames in the target speaker's representation.
In the decoding stage, the average of these top-$k$ target frames, ideally devoid of the original speaker identity but still conveying the speech content and emotion, is passed to a HiFi-GAN vocoder~\cite{kong2020hifi}, which synthesizes anonymized speech that retains the linguistic content of the source while adopting the vocal characteristics of the target speaker.

However, these systems have been shown to provide poor speaker privacy protection when facing stronger attacks~\cite{das2024comparing, xinyuan24_spsc}.
One possible reason is that SSL features are known to encode both content and speaker information. Consequently, although KNN retrieves the most similar frames from the target speaker, aiming to remove the timbre of the source speaker, the retrieved frames may still preserve speaker-specific cues to reflect the source speaker’s habitual speaking style, thereby leaking information about the source speaker under stronger attacks~\cite{das2024comparing, xinyuan24_spsc}. A variant of the KNN-based model~\cite{das2025} extends the previously described KNN approach by introducing two interpretable components that anonymize the duration and variation of phonemes to enhance privacy. However, this improvement in privacy comes at the cost of reduced utility.

\subsection{Proposed SEF-MK VAS}
As explained in Section~\ref{sec:knn}, KNN-based speaker-embedding-free methods have not utilized k-means, despite its proven ability to suppress speaker identity~\cite{miao2023language}. While k-means has been applied in disentanglement-based approaches, prior work typically uses a single k-means model trained on the full dataset and focuses mainly on the number of centroids, without thoroughly exploring its role in voice anonymization. To address this gap, we propose a novel speaker-embedding-free framework that employs multiple k-means quantizers, as shown in Figure~\ref{fig:overall}.

In the encoding stage, the input speech waveform $X$ is processed by WavLM\cite{chen2022wavlm} to extract SSL representations $h(X) = \mathbf{H} = [\mathbf{h}_1,...,\mathbf{h}_T]$, where each $\mathbf{h}_t \in \mathbb{R}^{1024}$ captures comprehensive speech characteristics, and $T$ is the number of frames in the input utterance. These SSL-based features inherently encode linguistic content, speaker identity, and emotional state~\cite{miao2023language, yang21c_interspeech}.

In the anonymization stage, the goal is to transform the SSL features to suppress speaker identity while preserving linguistic content and emotional expression. We observe that existing approaches~\cite{miao2023language} typically train a single k-means model on the entire data with a large set of speakers. However, because this clustering captures both phonetic and inter-speaker variations, the resulting centroids can unintentionally encode speaker-specific information, potentially leaking source speaker identity.
This raises an important but underexplored question: \textit{Does the composition of the k-means training data affect anonymization performance?} 
Intuitively, when k-means is trained on a single speaker’s data, the clustering focuses solely on intra-speaker phonetic variation, inherently avoiding the encoding of speaker identity. However, applying such a speaker-specific quantizer to features from a different speaker may introduce a mismatch that degrade the linguistic content. Using multiple k-means models may either improve the the extraction of linguistic content (i.e., better utility) or leaking speaker information (i.e., worse privacy).

To verify the above assumptions, the anonymization stage introduces a pool of $N$ specialized k-means models $\{KM_1,\ldots,KM_N\}$, where each model is trained on a distinct subset of speakers to encourage diverse clustering behavior. Each model uses $K = 1024$ clusters. During inference, a model $KM_n$ is randomly selected from the pool and applied to the SSL features $\mathbf{H}$ to assign each frame to a cluster, producing discrete assignments $\mathbf{Z} = [z_1,\ldots,z_T]$ with $z_t \in \{1,\ldots,K\}$, $\forall t \in \{ 1,\ldots,T\}$. Each assignment $z_t$ is then mapped to its corresponding cluster center embedding vector, yielding $\mathbf{E} = [\mathbf{e}_1,\ldots,\mathbf{e}_T]$, where $\mathbf{e}_t \in \mathbb{R}^{1024}$ represents the $1024$-dimensional centroid of cluster $z_t$. This operation is denoted as $\mathbf{E} = \text{Center}(\mathbf{Z})$.

We explore several strategies for constructing the pool of k-means models based on different speaker groupings in the training dataset. Let the dataset $D$ contain speech from $S$ speakers, and let $L$ be the number of speakers per group, with $L < S$. We define the following strategies:
\begin{itemize}
    \item \textbf{$D\text{-all}$}: All utterances from all $S$ speakers are used to train a single k-means model.

\item \textbf{$D\text{-}L\text{-sep}$}: The dataset is partitioned into groups of $L$ speakers, with each group used to train a separate k-means model, resulting in $\left\lfloor \frac{S}{L} \right\rfloor$ number of k-means models. Note that $D\text{-}1\text{-sep}$ denotes training a single k-means model for each speaker.  %resulting in $S / L$ models.
\end{itemize}
In our experiments, we evaluate the effects of different datasets and speaker group sizes. Specifically, we use either LibriSpeech-train-clean-460\footnote{\textit{LibriSpeech-train-clean-360} and \textit{LibriSpeech-train-clean-100} datasets~\cite{panayotov2015librispeech}} with 1,172 speakers ($S=1,172$) or VoxCeleb2~\cite{chung2018voxceleb2} with 5,994 speakers ($S=5,994$) as the dataset for $D$, and we conduct experiments using speaker group sizes $L \in \{1, 10, 20\}$.

The decoding stage reconstructs natural speech from the anonymized embeddings $E$ using a Conformer-based sequence model followed by a HiFi-GAN vocoder. The Conformer architecture combines self-attention mechanisms for capturing long-range dependencies with convolutional operations for local pattern modeling, transforming the input embeddings $\mathbf{E}$ into continuous representations $\mathbf{C} = [\mathbf{c}_1,\ldots,\mathbf{c}_T]$, where each $\mathbf{c}_t \in \mathbb{R}^{1024}$ corresponds to a 1024-dimensional frame-level feature vector. 
Note that during Conformer training, instead of using a single general k-means model trained on the entire training set, we use the matched k-means model trained on the same speaker as the input feature generator for the Conformer, aiming to best resynthesize the original speech.

We utilize the discrete representations (i.e., centroids of k-means models) as the input and the continuous representation as the target for the Conformer. The HiFi-GAN vocoder subsequently converts these intermediate representations into waveform samples, completing the anonymization pipeline while preserving natural speech characteristics.

%==========================================
\begin{figure}[t]
\centering
\includegraphics[width=0.46\textwidth]{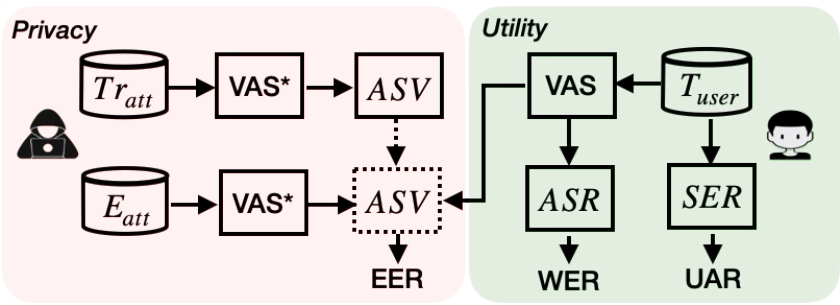}
\caption{Evaluation protocol following the VoicePrivacy 2024 guidelines. The subscript  \textit{att} means attacker. The databases are $Tr_{att}$: \textit{libri-train-360}; $E_{att}$: \textit{libri-dev-enroll} and \textit{libri-test-enroll}; $T_{user}$: \textit{libri-dev-trial-f}, \textit{libri-dev-trial-m}, \textit{libri-test-trial-f}, \textit{libri-test-trial-m}, \textit{IEMOCAP-dev}, and \textit{IEMOCAP-test}. If $\text{VAS}^* = \text{VAS}$, this represents a full attacker; if $\text{VAS}^* \cap \text{VAS} \neq \emptyset$, this represents a semi-attacker with partial knowledge of the user's VAS.}

\label{fig:protocol}
\vspace{-10pt}
\end{figure}   
%==========================================

\section{Experiments}
\subsection{Evaluation Protocol}

We follow the VPC 2024 evaluation protocol \cite{tomashenko2024voiceprivacy} to assess the effectiveness of our proposed VAS as plotted in Figure~\ref{fig:protocol}.

\subsubsection{Privacy Evaluation}
The attackers are assumed to have access to a few original or anonymized utterances for each speaker, referred to as \textit{enrollment} utterances and denoted as $E_{\text{att}}$, as well as some knowledge of the VAS. In the case of \textbf{SEF-MK}, the attackers are assumed to use a different k-means pool from the one used by the users during anonymization. This setup is referred to as a \textit{semi-attacker}. If the attacker and the user share the same k-means pool, we refer to it as a \textit{full attacker} scenario. Note that the VPC 2024 focuses on the \textit{semi-attacker} scenario. In the following experiments, we also examine the \textit{full attacker} scenario from the user’s point of view to understand the worst-case situation.\footnote{It is useful to measure system performance in the worst-cast situation, but it is unlikely to happen in applications if the users or vendors use a k-means pool constructed upon undisclosed data that the attacker cannot access.}

The attackers employ an ASV model, specifically, the ECAPA-TDNN model~\cite{desplanques2020ecapa}, trained on anonymized speech, denoted as $Tr_{\text{att}}$, to reduce the mismatch between original and anonymized utterances and infer the speaker’s identity. The Equal Error Rate (EER) is used to evaluate the privacy protection capability of the VAS, with an EER close to 50\% indicating perfect privacy protection.

\subsubsection{Utility Evaluation}

The utility evaluation depends on the downstream tasks. We follow VPC 2024 and consider ASR and emotion analysis. 
The evaluation of speech content and emotion preservation of anonymized \emph{test trial} speech, denoted as $T_{user}$, is straightforward.
The speech content preservation ability in anonymized speech is assessed by the word error rate (WER) computed using an ASR evaluation model\footnote{\url{https://huggingface.co/speechbrain/asr-wav2vec2-librispeech}}. A lower WER, similar to that of the original speech, indicates good speech content preservation ability.
The emotion preservation ability in anonymized speech is assessed by unweighted average recall (UAR) produced by a pre-trained speech emotion recognizer (SER), 
which is a wav2vec2-based system~\cite{tomashenko2024voiceprivacy}. A higher UAR, similar to that of the original speech, indicates better emotional content preservation ability.

\subsection{Datasets and System Configurations}

The \textbf{SEF-MK} VAS is built using the publicly available WavLM-large\footnote{\url{https://huggingface.co/microsoft/wavlm-large}}. 
The Conformer encoder consists of 6 layers with a kernel size of 31, each with 4 attention heads, and an MLP with a hidden dimension of 2048, trained on \textit{LibriSpeech-train-clean-460}~\cite{panayotov2015librispeech}.
We directly utilizes the HiFi-GAN provided in~\cite{knn_vc}.
This training setup is compatible with the VPC 2024 evaluation protocol.
VASs are evaluated on the official VPC 2024 dev and test sets \cite{tomashenko2022voiceprivacy}.
It contains English utterances by several female and male speakers from the \textit{LibriSpeech} corpora, split into dev and test sets.

%==========================
\begin{table}[t]
\caption{Results (in \%) for various k-means training strategies applied to the user on development sets under the full-attacker scenario. \textbf{Libri-all} denotes a single k-means model trained on the entire dataset. \textbf{D-L-sep} denotes training one k-means model per \textbf{L} speakers. \textbf{Libri} refers to the \textit{LibriSpeech-train-clean-460} dataset with 1,172 speakers, and \textbf{Vox} refers to the \textit{VoxCeleb2} dev dataset with 5,994 speakers.}
\label{tab:kmeans-strategy}
\centering
\footnotesize
\begin{tabular}{lllcccc}  
\toprule
 & &  \multicolumn{2}{c}{{EER} $\uparrow$}  & {WER} $\downarrow$ & {UAR} $\uparrow$ \\  
\cmidrule(lr){3-4} \cmidrule(lr){5-5} \cmidrule(lr){6-6}
&  \#k-means & dev-f & dev-m & dev & dev \\
\cmidrule(lr){3-4} \cmidrule(lr){5-5} \cmidrule(lr){6-6}
Original & - & 10.51 & 0.93 & 1.80 & 69.08 \\
Resyn   & - & 17.33 & 6.55 & 3.47 & 59.99 \\
\midrule
Libri-all & 1 & 22.59 & 13.34 & 6.06 & 49.94 \\
Libri-20-sep & 58 & \textbf{23.86} & 11.36 & \textbf{3.33} & \textbf{56.89} \\
Libri-10-sep & 117 & 21.13 & 11.80 & 3.37 & 55.50 \\
Libri-1-sep & 1,172 & 21.71 & {15.84} & 3.99 & 48.01 \\
\midrule
Vox-all & 1 & 19.04 & \textbf{16.00} & 4.98 & 48.99 \\
Vox-20-sep & 299 & 18.89 & 14.29 & 4.58 & 49.23 \\
Vox-10-sep & 599 & 19.32 & 12.27 & 4.65 & 48.66 \\
Vox-1-sep & 5,994 & 22.73 & 15.96 & 7.64 & 45.67 \\
\bottomrule 
\end{tabular}
\vspace{-10pt}
\end{table}

\begin{table}[t]
\caption{EER results (\%) for various k-means training strategies applied to the attacker on test and development sets under the \textit{semi-attacker} scenario.}
\label{tab:attack}
\centering
\footnotesize
\begin{tabular}{lllcccc}  
\toprule
 & & \multicolumn{4}{c}{{EER (\%)} $\uparrow$} \\  
\cmidrule(lr){3-6}
 & \#k-means & dev-f & dev-m & test-f & test-m \\
\cmidrule(lr){3-6}
Original & & 10.51 & 0.93 & 8.76 & 0.42 \\
\midrule
\multicolumn{6}{l}{\textit{User uses 58 k-means models: Libri-20-sep}} \\
\midrule
Libri-all & 1 & 42.90 & 35.22 & 38.53 & 35.38 \\
Vox-all  & 1 & 28.95 & 25.48 & 27.19 & 28.06 \\
Vox-1-sep  & 5,994 & 25.98 & 14.44 & 15.16 & 15.59 \\
\midrule
\multicolumn{6}{l}{\textit{User uses 1,172 k-means models: Libri-1-sep}} \\
\midrule
Libri-all & 1 & 44.62 & 41.17 & 42.93 & 39.68 \\
Vox-all & 1 & 37.20 & 37.57 & 38.14 & 37.17 \\
Vox-1-sep  & 5,994 & 26.56 & 19.56 & 16.42 & 17.60 \\

\midrule
\multicolumn{6}{l}{\textit{User uses one k-means model: Libri-all}} \\
\midrule
Vox-all  & 1 & 34.20 & 36.01 & 37.03 & 37.39 \\
Libri-1-sep  & 1,172 & 23.31 & 16.62 & 16.44 & 17.87 \\
Vox-1-sep  & 5,994 & 26.14 & 20.34 & 18.40 & 19.12 \\
\bottomrule 
\end{tabular}
\vspace{-10pt}
\end{table}
%==========================

%==========================
\begin{table*}[t]
\caption{Results (\%) on various VASs. B1-B6 denotes the six baseline models in VPC 2024.}
\label{tab:baseline-compare}
\centering
\footnotesize
\begin{tabular}{lllcccccccccc}  
\toprule
 & &  \multicolumn{5}{c}{EER $\uparrow$} & \multicolumn{3}{c}{WER $\downarrow$} &  \multicolumn{3}{c}{UAR $\uparrow$} \\  
\cmidrule(lr){3-7} \cmidrule(lr){8-10} \cmidrule(lr){11-13}
& & dev-f & dev-m & test-f & test-m &  \cellcolor{gray!7}avg &  dev & test & \cellcolor{gray!7}avg & dev & test &  \cellcolor{gray!7}avg \\
\cmidrule(lr){3-7} \cmidrule(lr){8-10} \cmidrule(lr){11-13}
& Original  & 10.51 & 0.93 & 8.76 & 0.42 & \cellcolor{gray!7}5.66 & 1.80 & 1.85 & \cellcolor{gray!7}1.83 & 69.08& 71.06&  \cellcolor{gray!7}70.07 \\
\midrule
\multirow{5}{*}{\rotatebox{90}{Disentangle}}  & B1~\cite{b1}  &  10.94 & 7.45 & 7.47 & 4.68 &  \cellcolor{gray!7}7.64    & 3.07 & 2.91  &  \cellcolor{gray!7}2.99 & 42.71 & 42.78 &   \cellcolor{gray!7}42.75 \\ 
& B3~\cite{b3}  &  28.43 & 22.04 & 27.92 & 26.72 &  \cellcolor{gray!7}26.78 & 4.29 &  4.35 &  \cellcolor{gray!7}4.32 & 38.09 & 37.57  & \cellcolor{gray!7}37.83 \\  
 & B4~\cite{b4}  &  34.37 & 31.06 & 29.37 & 31.16 &  \cellcolor{gray!7}31.99 & 6.15 & 5.90 & \cellcolor{gray!7}6.03  & 41.97 & 42.78 & \cellcolor{gray!7}42.38  \\

& B5~\cite{b5}  &  35.82 & 32.92 & 33.95 & 34.73 &  \cellcolor{gray!7}34.36  & 4.73 & 4.37  &  \cellcolor{gray!7}4.55 &38.08 & 38.17 & \cellcolor{gray!7}38.13 \\ 
& B6~\cite{b5}  &   25.14 & 20.96 & 21.15 & 21.14 &  \cellcolor{gray!7}22.10  & 9.69 & 9.09  &  \cellcolor{gray!7}9.39 & 36.39 & 36.13 &  \cellcolor{gray!7}36.26\\
& OH~\cite{miao2025adapting} & 44.89 & 34.74 &  \cellcolor{gray!7}39.26& 37.64 &38.54 &2.36& 2.48 & \cellcolor{gray!7}2.42 & 47.01 &47.37 &  \cellcolor{gray!7}47.19 \\
\midrule
DSP & B2~\cite{b2}  &  12.91 & 2.05 & 7.48 & 1.56 &  \cellcolor{gray!7}6.00   & 10.44 & 9.95 &  \cellcolor{gray!7}10.20 & 55.61 
& 53.49  & \cellcolor{gray!7}54.55  \\ 

\midrule
\multirow{3}{*}{\rotatebox{90}{SEF}} & KNN*~\cite{cai2024privacy} &  18.35 & 13.66 & 16.24 & 12.50 &  \cellcolor{gray!7}15.19  & {2.99} & {2.96} & \cellcolor{gray!7}\textbf{2.98} & 47.70 & 50.61 &  \cellcolor{gray!7}49.16 \\
& Private KNN*~\cite{das2025} &  - & - & - & -&  \cellcolor{gray!7}\textbf{49.40}  & - & -&  \cellcolor{gray!7}4.80 & - & - &  \cellcolor{gray!7}49.40 \\ 
& SEF-MK & 42.90  & 35.22 & 38.53 & 35.38 &  \cellcolor{gray!7}38.01  &  3.33 &  3.30 & \cellcolor{gray!7}3.31 & {56.89} & {58.31} & \cellcolor{gray!7}\textbf{57.60} \\
\bottomrule 
\multicolumn{13}{c}{*Note that these systems use the same target speaker pool for both the user and the attacker, more likely a \textit{full attacker} scenario.}
\vspace{-5mm}
\end{tabular}
\end{table*}
%==========================

\section{Results and Discussion}
In this section, we begin from the user's point of view to find the best configurations for the proposed \textbf{SEF-MK}, focusing on how the composition of the k-means training data affects performance under the \textit{full-attacker} scenario, where the attacker shares the same k-means pool as the user, i.e., a worst-case (but unrealistic) setting. After determining the optimal user settings, we shift to the attacker's point of view to investigate how the attacker can exploit the composition of the k-means training data. When the attacker isolates the k-means pool, this corresponds to evaluating \textbf{SEF-MK} under the standard VPC scenario, i.e., the \textit{semi-attacker} setting. Finally, we compare the results with those of other VASs under both fully and semi-attacking scenarios.

\subsection{How can users configure {SEF-MK} under the \textit{full-attacker} scenario}
\label{sec:exp:fullattacker}

Under this scenario, assuming the worst case where the VAS is fully shared with the attacker, the user aims to find a configuration that maximizes privacy (high EER) while maintaining good utility (low WER and high UAR).

We experiment with different k-means training partition strategies on the development sets, with the results summarized in Table~\ref{tab:kmeans-strategy}. At the top of the table, we evaluate the performance of the resynthesized \textbf{SEF-MK}, denoted as `resyn', where each utterance uses its own data to train the corresponding k-means model and generate speech. Performance that closely matches the original speech indicates better generation quality. For `resyn', the EER is close to that of the original speech, while both WER and UAR show reasonable degradations, reflecting the trade-offs introduced by the waveform resynthesis process. This result is consistent with those observed on other mainstream anonymization systems~\cite{panariello23_spsc}.

The middle and bottom sections of the table evaluate different training partition strategies using two datasets: the smaller \textit{LibriSpeech-train-clean-460} dataset with 1,172 speakers, and the larger \textit{VoxCeleb2} dev dataset with 5,994 speakers.
For both datasets, we did not observe significant changes in EER when switching from a single k-means model (Libri-all, Vox-all) to multiple models (Libri-20/10/1-sep, Vox-20/10/1-sep). However, WER and UAR consistently improve when using multiple k-means models. The best WER and UAR results are achieved when the k-means models are trained on data from 20 speakers per model, for both the Libri and Vox datasets.

When comparing results between the Libri and Vox datasets under the same configuration, we observe that using the larger dataset (Vox) slightly increases the EER, particularly for male speech, but the utility metrics drop more significantly. One potential reason could be that the VoxCeleb2 data, which was sourced from Celebrities' interviews in the wild, is considerably more noisy than the audiobook-based Librispeech data. Relatively clean data may be preferred for building the k-means models that can encode the linguistic contents. Overall, the Libri-20-sep configuration provides a better trade-off between privacy and utility.

\subsection{How can attacker configure {SEF-MK} under the \textit{semi-attacker} scenario}

After exploring the configuration of \textbf{SEF-MK} from the user's perspective, it is also important to examine the attacker's configuration. The attacker's goal is to use a similar VAS to anonymize speech in a way that mimics the user's approach, optimize the ASV model trained on anonymized speech, and attempt to trace the original speaker identity from the anonymized speech. This only impacts the privacy aspect of the system. Hence, Table~\ref{tab:attack} presents the EER results on both development and test sets for various attacker configurations, assuming fixed user settings.

We first examine the attackers' performance, assuming users using `Libri-20-sep' (the best configuration for users as reported in the section~\ref{sec:exp:fullattacker}). 
At the top of the table, we list various attacker configurations. Interestingly, we observe that when the attacker uses only a single k-means model to generate anonymized speech, the resulting EER is significantly higher, e.g., over 35\% for `Libri-all' and over 25\% for `Vox-all', indicating poor attacking effectiveness (and good privacy protection from the user's point of view). However, when the attacker employs multiple k-means models, even with different training data (e.g., Vox-1-sep), the attack becomes much more effective, reducing the EER to around 15\%. We confirm this trend by testing under different user settings, as shown in the middle and bottom sections of the table, and consistently observe the same pattern.

One possible reason is that using randomly selected multiple k-means models produces more diverse anonymized utterances compared to using a single k-means model, which benefits the attacker. The attacker's ASV model trained on more diverse anonymized utterances may perform better in discriminating incoming anonymized utterances and exploit more speaker-specific attributes that are not fully removed by \textbf{SEF-MK}. 

\subsection{Comparasion with other VASs}

Finally, we select the best semi-attacker configuration of our system where the user uses `libri-20-sep', and the attacker uses `Libri-all' as the k-means pool, and compare it with other VASs, as shown in Table~\ref{tab:baseline-compare}. Among the disentanglement- and DSP-based VASs, speaker-embedding-free (SEF) KNN-based methods achieve decent performance even under full attacker scenarios. However, \textbf{SEF-MK} achieves the highest UAR while maintaining a good balance between privacy protection and content preservation.

Note that the private KNN system~\cite{das2025} obtained the highest EER likely because of the additional component to anonymize the duration of the phones, but it may also have affected the utility of the anonymized speech. The anonymization of the durations can also be incorporated in our proposed system, but this is left to future work.

\begin{figure}[t]
\centering
\scriptsize

\subfloat[][WavLM: $ \mathbf{H} $]{%
  \includegraphics[width=0.45\textwidth]{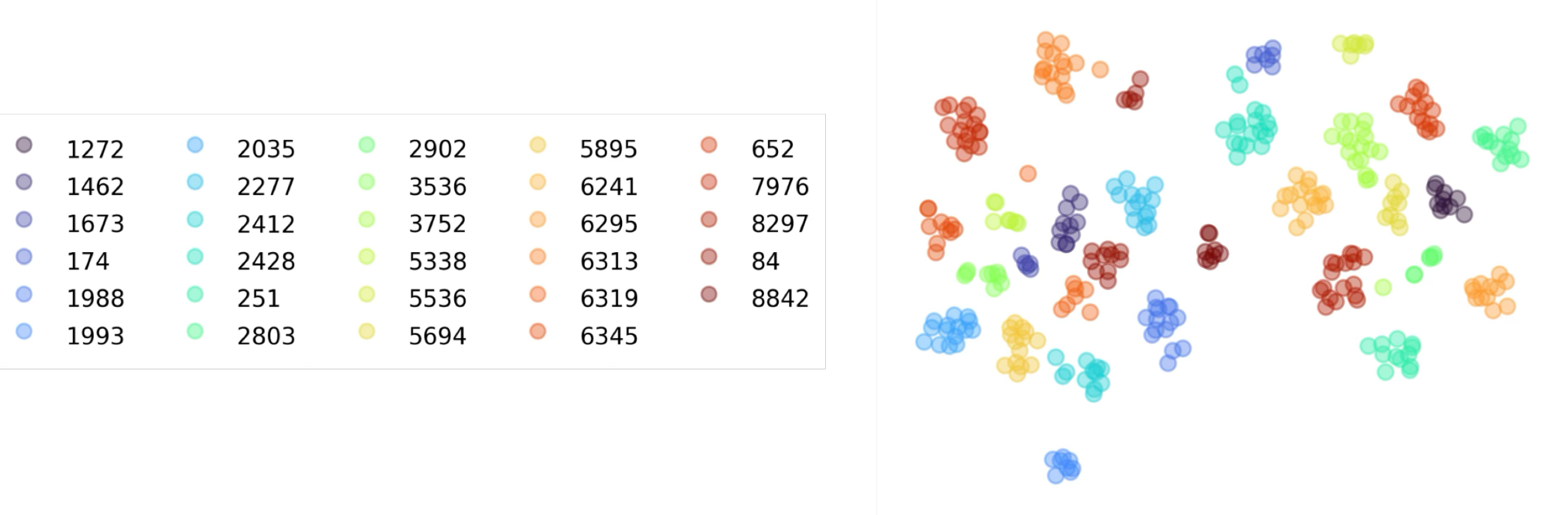}%
  \label{fig:wavlm}
} \\[0.1em]

\subfloat[][k-means: $E_{\text{libri-all}} $]{%
  \includegraphics[width=0.2\textwidth]{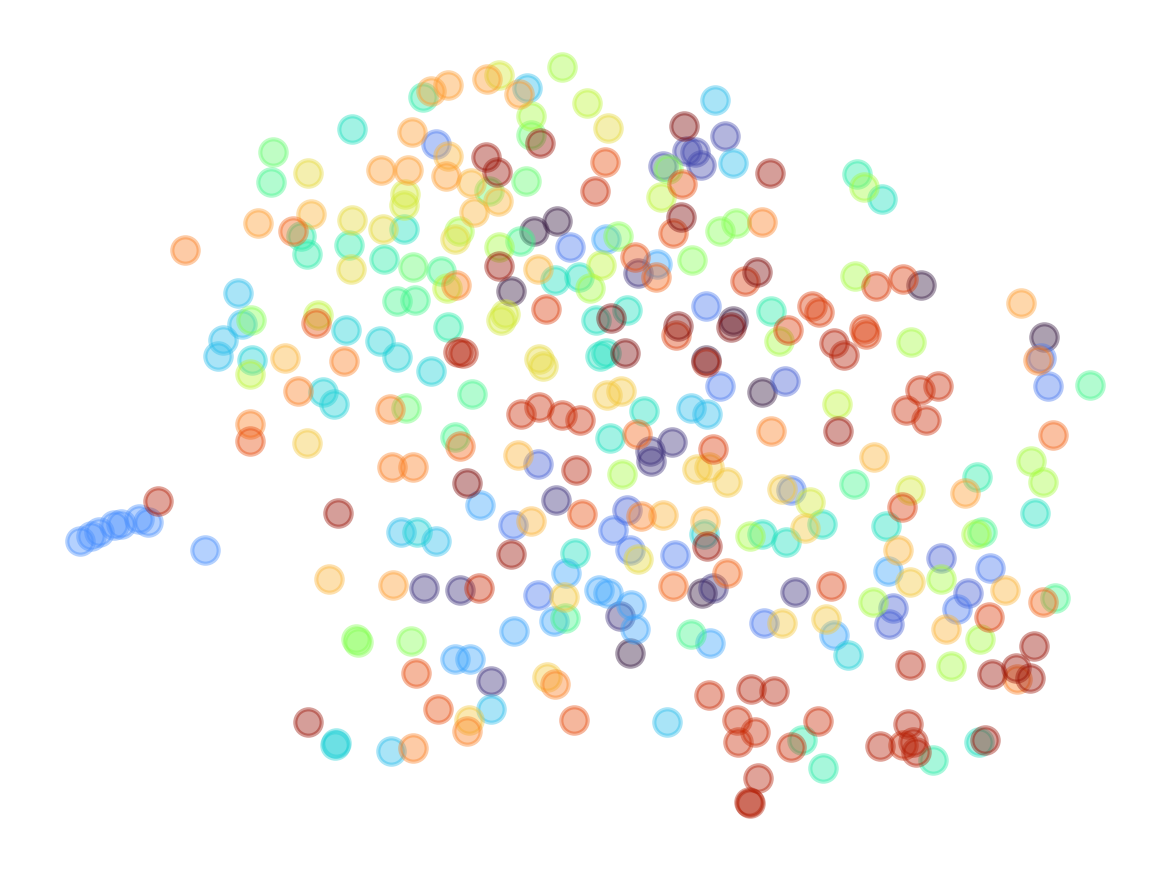}%
  \label{fig:km-spkall}
}
\quad
\subfloat[][k-means: $E_{\text{libri-1-sep}} $]{%
  \includegraphics[width=0.2\textwidth]{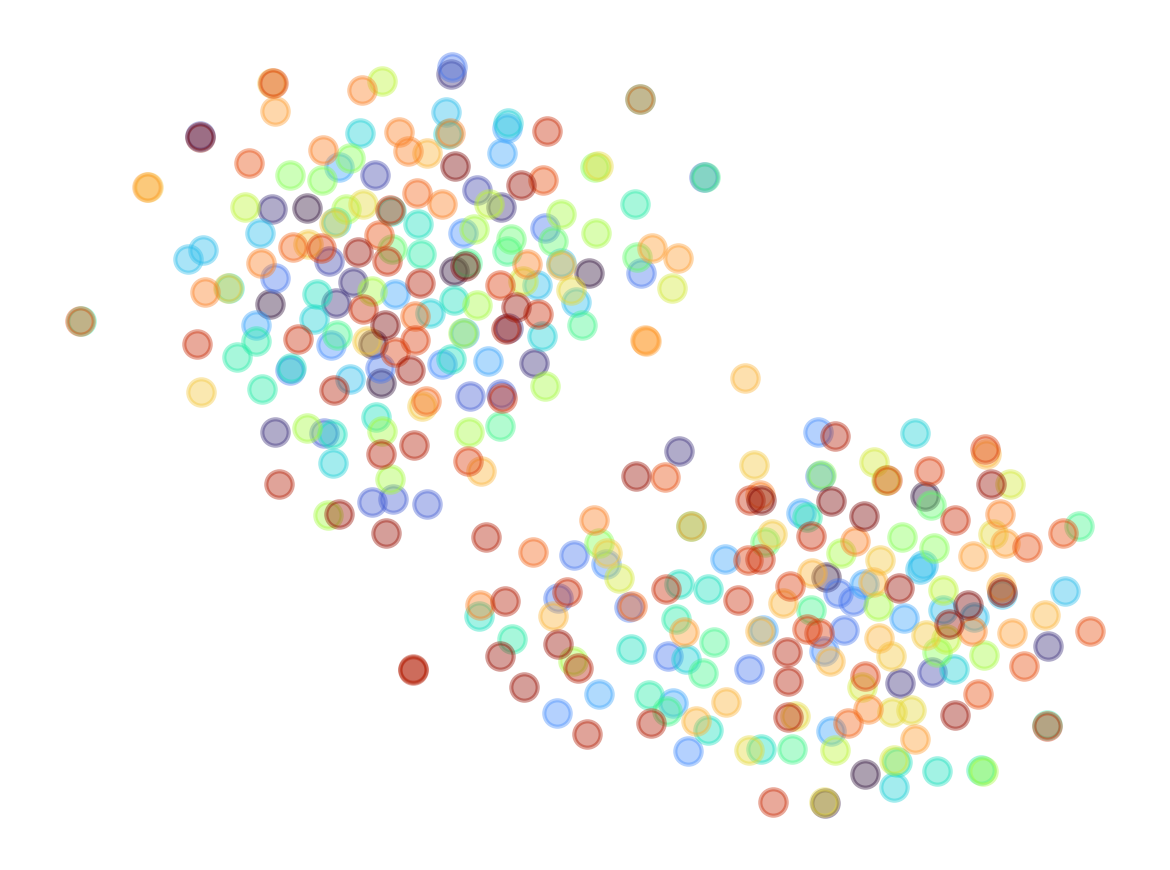}%
  \label{fig:km-sep}
} \\[0.1em]

\subfloat[][Conformer: $C_{\text{libri-all}} $]{%
  \includegraphics[width=0.2\textwidth]{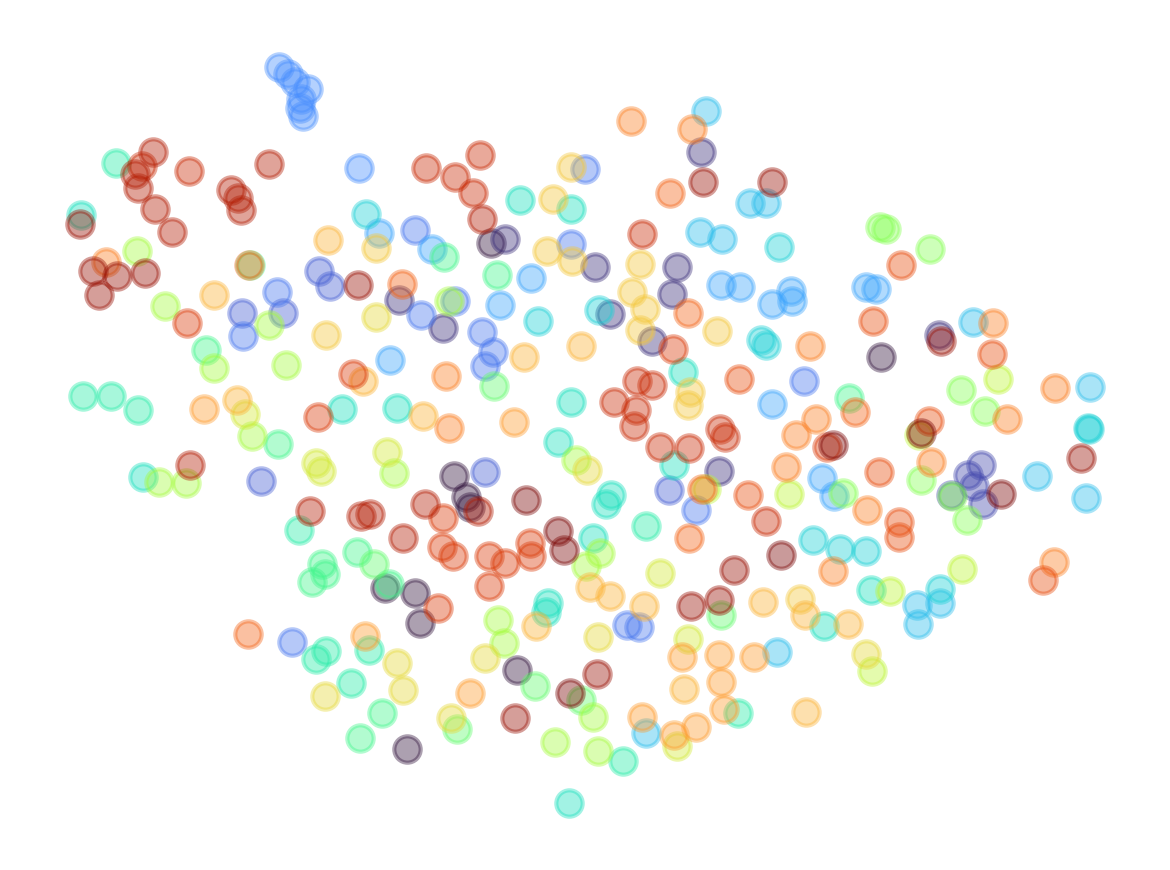}%
  \label{fig:con-spkall}
}
\quad
\subfloat[][Conformer: $C_{\text{libri-1-sep}} $]{%
  \includegraphics[width=0.2\textwidth]{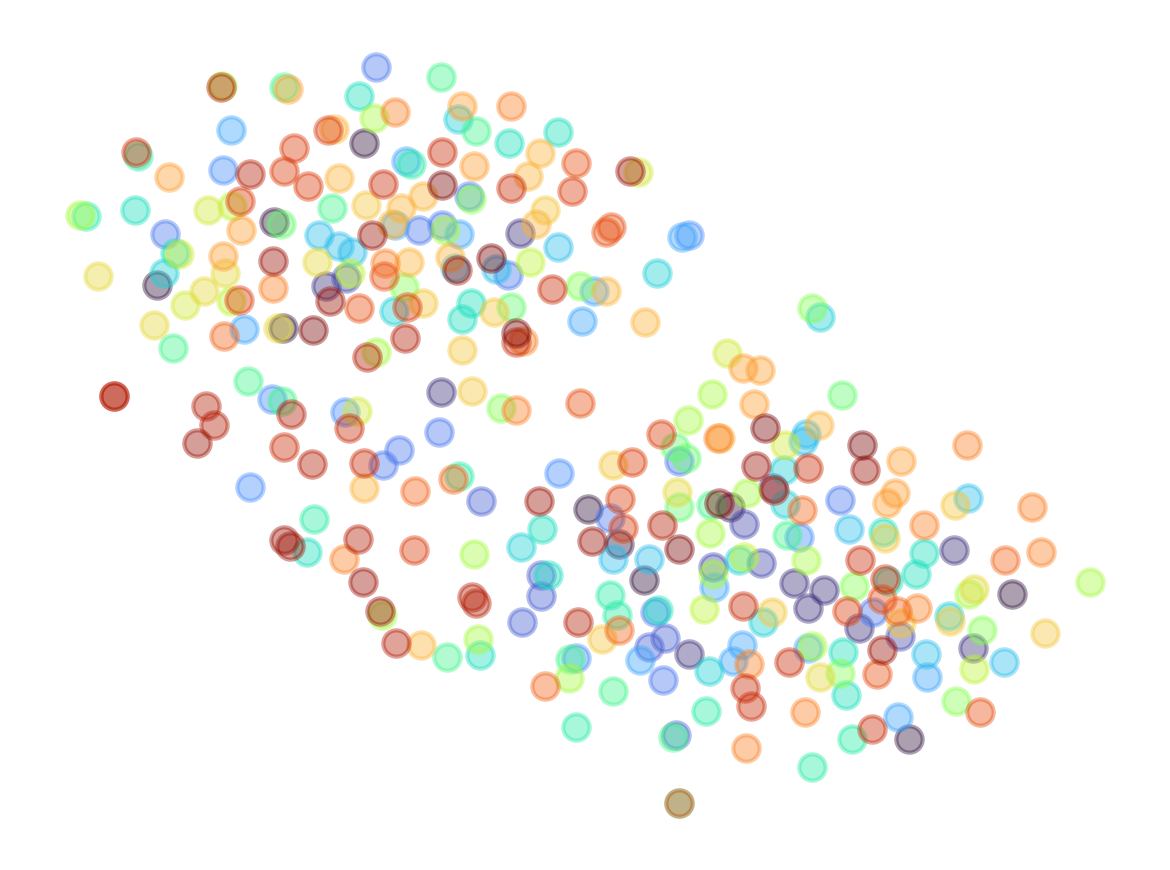}%
  \label{fig:con-sep}
}

\caption{\textit{Libri-dev-enroll:} Comparison of WavLM representations and clustering-based embeddings across speaker and separation settings. Different speaker identities are marked using different colors.}
\label{fig:raw-emb}

\vspace{-10pt}
\end{figure}

\subsection{Furthur analysis}
To further visualize the effectiveness of the proposed \textbf{SEF-MK}, we use t-SNE~\cite{van2008visualizing} to visualize the k-means and Conformer output embeddings for both the single-model setting (\textit{libri-all}) and the multi-k-means model setting (\textit{libri-1-sep}) as shown in Figure~\ref{fig:raw-emb}.
The speaker embeddings are extracted from 29 speakers in the \textit{libri-dev-enroll} dataset, with each speaker represented by a different color. In Figure~\ref{fig:raw-emb}(a), the WavLM features form clear and distinct speaker clusters, indicating that they encode a significant amount of speaker identity-related information. However, after applying k-means and the Conformer, these clusters become indistinct, and speaker separability is no longer clearly visible. Additionally, the embedding distributions before and after the Conformer remain similar, demonstrating its strong feature reconstruction capability. 
Furthermore, compared to the single k-means model, the multiple k-means models also result in embeddings with less speaker information, but they form two clusters (which probably represent the genders). This echoes with the results in Sec.~\ref{sec:exp:fullattacker} that \textit{libri-1-sep} degraded the privacy protection (i.e., higher EER) for one of the gender. 

\section{Conclusion}
In this work, we propose a speaker-embedding-free voice anonymization system that leverages multi-k-means quantization on SSL features. Specifically, our method utilizes a pool of k-means models, with one randomly selected at either the frame level or the utterance level during inference. We systematically analyze the impact of different k-means anonymization strategies from both the user's and the attacker's perspectives. Experimental results demonstrate that our approach achieves superior utility preservation while providing strong privacy protection, outperforming various existing VASs.

\bibliographystyle{IEEEtran}
\bibliography{main}
\end{document}